# Phonon heat conduction across slippery interfaces in twisted graphite

Fuwei Yang[1,2,3,4#], Wenjiang Zhou[1,5,6#], Zhibin Zhang[7#], Xuanyu Huang[3,4,8#], Jingwen Zhang[9], Nianjie Liang[1,5], Wujuan Yan[1,5], Yuxi Wang[1,5], Mingchao Ding[10], Quanlin Guo[7], Yu Han[4], Te-Huan Liu[9*], Kaihui Liu[7,11,12*], Quanshui Zheng[2,3,4,8,13*] and Bai Song[1,5*]

[1]National Key Laboratory of Advanced MicroNanoManufacture Technology, Peking University, Beijing, 100871, China.

[2]Department of Engineering Mechanics, Tsinghua University, Beijing, 100084, China.

[3]Center for Nano and Micro Mechanics, Tsinghua University, Beijing 100084, China.

[4]Institute of Superlubricity Technology, Research Institute of Tsinghua University in Shenzhen, Shenzhen 518057, China.

[5]College of Engineering, Peking University, Beijing 100871, China.

[6]School of Advanced Engineering, Great Bay University, Dongguan 523000, China.

[7]State Key Laboratory for Mesoscopic Physics, Frontiers Science Centre for Nano-optoelectronics, School of Physics, Peking University, Beijing 100871, China

[8]Institute of Materials Research, Shenzhen International Graduate School, Tsinghua University, Shenzhen, 518055, China.

[9]School of Energy and Power Engineering, Huazhong University of Science and Technology, Wuhan, Hubei 430074, China.

[10]Beijing National Laboratory for Condensed Matter Physics, Institute of Physics, Chinese Academy of Sciences, Beijing, China

[11]International Centre for Quantum Materials, Collaborative Innovation Centre of Quantum Matter, Peking University, Beijing 100871, China

[12]Songshan Lake Materials Laboratory, Institute of Physics, Chinese Academy of Sciences, Dongguan, China

[13]Center of Double Helix, Shenzhen International Graduate School, Tsinghua University, Shenzhen, 518055, China.

[#]These authors contributed equally to this work.

[*]Corresponding author. Email: songbai@pku.edu.cn (B.S.); zhengqs@tsinghua.edu.cn (Q.Z.); khliu@pku.edu.cn (K.L.); thliu@hust.edu.cn (T.H.L.)

**Abstract**

Interlayer rotation in van der Waals (vdW) materials offers great potential for manipulating phonon dynamics and heat flow in advanced electronics with ever higher compactness and power density[1]. However, despite extensive theoretical efforts in recent years[2–8], experimental measurements remain scarce especially due to the critical challenges of preparing single-crystalline twisted interfaces and probing interfacial thermal transport with sufficient resolution[9–12]. Here, we exploited the intrinsic twisted interfaces in highly oriented pyrolytic graphite (HOPG). By developing novel experimental schemes based on microfabricated mesas, we managed to achieve simultaneous mechanical characterizations and thermal measurements. In particular, we pushed the HOPG mesas with a microprobe to identify and rotate single-crystalline intrinsic interfaces owing to their slippery nature as is well known in structural superlubricity[13–16]. Remarkably, we observed over 30-fold suppression of thermal conductance for the slippery interfaces by using epitaxial graphite as a control. Nonetheless, the interfacial conductance remains around 600 MWm$^{-2}$K$^{-1}$ which surpasses the highest values for artificially stacked vdW structures by more than five times[12,17,18]. Further, atomic simulations revealed the predominant role of the transverse acoustic phonons. Together, our findings highlight a general physical picture that directly correlates interfacial thermal transport with sliding resistance, and lay the foundation for twist-enabled thermal management which are particularly beneficial to twistronics[19,20] and slidetronics[21–24].

Twisted van der Waals (vdW) materials have recently emerged as a unique platform for exploring a vast number of exotic phenomena[25,26] such as unconventional superconductivity[27], quantum anomalous Hall effect[28,29], topological polaritons[30], and nanolasing[31]. Of central importance is the delicate manipulation of electronic structures through the moiré superlattice. Meanwhile, phonon dynamics and transport in the moiré superlattice have also garnered considerable attention[5,32–37]. In particular, by stacking a few layers of polycrystalline molybdenum disulfide with random interlayer rotations, extremely anisotropic heat conductors have been demonstrated for advanced thermal management of electronics[38]. However, despite the great fundamental and applied interest in heat flow across twisted interfaces, most studies are limited to theoretical calculations[2–8] while experimental measurements are rare due to a range of critical challenges, especially for high-quality single-crystalline interfaces[10–12].

To begin with, artificially stacked vdW structures often suffer from wrinkles and contaminants, which could potentially smear out twist-induced variations or introduce unexpected artifacts. Even though direct growth of twisted bilayers has recently been achieved[39,40], measurement of interfacial thermal transport in atomically thin membranes still presents a notable obstacle due to complications from their substrate and various contact resistances. Alternatively, highly oriented pyrolytic graphite (HOPG, see Fig. 1a for representative optical images) offers many intrinsic twisted interfaces, which appears unique among diverse vdW materials. These interfaces feature ultralow friction and therefore have been widely employed in the field of structural superlubricity[13–15], both for the fundamental understanding of mechanical friction[41,42] and for various emerging devices that demand mechanical movements[19–21], such as dynamic Schottky generators[22] and sliding electrical contacts[24]. In addition, this system has also enabled the study of angle-dependent electron transport[43]. However, unlike mechanical and electrical properties, the thermal conductance of a single buried interface in HOPG remains challenging to measure. Although the collective thermal resistance of multiple slippery interfaces can be measured, the resulting physical insight is limited since the twist angles are uncontrolled and unknown.

To overcome these challenges and understand the effect of intrinsic twisted interfaces on heat transport, we first developed novel experimental schemes based on microfabricated graphite mesas. These mesas not only substantially increase the sensitivity of interfacial thermal measurement, but also enable simultaneous mechanical characterizations which prove key to revealing the underlying mechanisms. In addition, we took advantage of epitaxially grown single-crystal graphite[44] (EG, Fig. 1b) which is free of twisted interfaces and therefore can serve as an excellent reference to compare with HOPG. This is schematically illustrated in Fig. 1e,f, and experimentally demonstrated in the cross-sectional scanning transmission electron microscopy (STEM) images in Fig. 1c,d, with many intrinsic interfaces in HOPG clearly observed (marked with white dashed lines) and their typical distances labeled. The insets in Fig. 1c,d further show images of individually resolved graphene layers.

Representative optical images of large arrays of HOPG and EG mesas are shown in Fig. 1g,h. For systematic analysis, all these circular mesas were fabricated together (Methods), with heights around 1 μm and nominal radii from 2 μm to 5 μm. Closer inspections with scanning electron microscopy (SEM) are presented in Fig. 1i,j. Based on these mesas, we further demonstrate the structural difference between HOPG and EG over larger areas via mechanical characterizations. As illustrated in Fig. 1o, we used a tungsten microprobe to shear the mesa at a given normal load, and recorded the lateral force when the mesa top was moved and a new interface was exposed [Methods and Supplementary Information (SI) including Note 9 and Video 1]. We repeated this experiment for many mesas of similar size at different positions on HOPG and EG, and found that the lateral force for sliding the EG mesas was indeed two orders of magnitude larger (Fig. 1o).

After shearing, we inspected the exposed interfaces by optical (Fig. 1k,l) and atomic force microscopy (AFM, Fig. 1m,n), and Raman spectroscopy (Supplementary Fig. 24). Notably, multiple layers of the EG mesas were simultaneously sheared due to similarly high interlayer sliding resistances. In comparison, only one single-crystalline interface was exposed for HOPG mesas, as evident from its slippery nature[13,16].

For thermal transport measurement, we employed the laser pump-probe technique of frequency-domain thermoreflectance (FDTR) with the graphite samples coated by a gold (Au) nanofilm transducer (Methods and SI). For preliminary study, we focus on bulk samples. The ultrahigh anisotropy (~300) of graphite presents a major challenge (Supplementary Note 4), since a small uncertainty in the in-plane thermal conductivity ($k_i$) can lead to a large error in the out-of-plane conductivity ($k_o$). To alleviate this problem, we combined the beam-offset method (Fig. 1p inset) with single-point frequency sweep ($f$-sweep, Fig. 1q inset) and simultaneously extracted $k_i$ and $k_o$ via iterative fitting (Supplementary Fig. 10). For direct comparison, the HOPG and EG samples were always Au-coated together (Supplementary Notes 1 and 2). As shown in Fig. 1p,q, the representative FDTR phases for EG are consistently higher than the corresponding signals for HOPG in both the beam-offset and $f$-sweep measurements, which intuitively indicate better thermal spreading capability of the former[45]. Indeed, the fitted $k_i$ and $k_o$ for EG at room temperature (RT, 295 K) are 1980 ± 130 and 6.9 ± 2.2 Wm$^{-1}$K$^{-1}$, respectively, both higher than the values for HOPG (1860 ± 120 and 5.5 ± 1.8 Wm$^{-1}$K$^{-1}$). Unfortunately, the uncertainty of $k_o$ remains too large to warrant a reliable comparison (Supplementary Fig. 13), despite our effort to go beyond standard FDTR approaches.

To substantially increase the accuracy of out-of-plane measurement, we took advantage of the graphite mesas and adapted the FDTR technique accordingly by incorporating finite-element method (FEM) into the fitting process (Methods). The idea is to create a scenario where in-plane transport is truncated by the edge of the mesa so that heat predominantly flows in the out-of-plane direction (Fig. 2a). As a result, the sensitive parameters change from $k_i$ and the laser spot sizes to $k_o$ and the geometric parameters of the mesa which can be accurately determined (Supplementary Note 7 and Fig. 22). As shown in Fig. 2b, a typical graphite mesa was first identified from FDTR mapping, and then single-point $f$-sweep measurements were performed at the center of the mesa. In Fig. 2c, we demonstrate the FEM-simulated effective temperature rise in a mesa-based FDTR measurement at four representative pump modulation frequencies $f$. It is evident that thermal transport in the graphite mesa is essentially one-dimensional especially at low frequencies, and we therefore focus on relatively small $f$ in subsequent discussions. Representative FDTR phases for HOPG mesas with different radii ($r$) and EG mesas measured at different temperatures are plotted in Fig. 2d,e, respectively (additional data in Supplementary Fig. 14). Excellent fitting is achieved in all cases, confirming the reliability of this method (Supplementary Note 6).

In Fig. 2f, we present the out-of-plane thermal conductivity measured at room temperature from graphite mesas of varying sizes, with at least ten different samples at each radius. The values for the EG mesas are generally higher than those of the HOPG mesas with similar radii. Interestingly, $k_o$ exhibits a rising trend with increasing mesa radius for both HOPG and EG, reaching respectively 6.4 ± 0.5 Wm$^{-1}$K$^{-1}$ and 6.9 ± 0.5 Wm$^{-1}$K$^{-1}$ for the largest mesas with $r \approx 5$ μm. This unexpected size effect can be understood by including the contribution from the mesa edge where a thin amorphous region forms during the etching process[46]. Considering heat flow through both the edge and interior of the mesa in a parallel transport model (Supplementary Note 10), $k_o$ can be written as $k_o = k_{int} - 2(k_{int} - k_{edge})t/r$. Here, $k_{edge}$ and $t$ are the thermal conductivity and thickness of the edge, respectively, while $k_{int}$ is the thermal conductivity of the interior which can be approached at sufficiently large $r$ and therefore represents the bulk limit. By fitting the room-temperature experimental data, we obtained $k_{int}$ = 6.6 Wm$^{-1}$K$^{-1}$ and 7.4 Wm$^{-1}$K$^{-1}$ for HOPG and EG, respectively. With the 5 μm-mesas, we further performed experiments at varying temperatures. As shown in Fig. 2g, the out-of-plane thermal conductivity of EG is again consistently higher than that of HOPG. Notably, the error bars are much smaller than the results from bulk graphite (Supplementary Fig. 13), and can finally allow us to extract the thermal conductance of twisted interfaces based on the different $k_o$ of HOPG and EG.

To begin with, we assume the same thermal conductance for all the slippery interfaces in HOPG despite their unknown and different twist angles. This assumption is supported by previous calculations for both interfacial friction and thermal transport that show notable angle dependences only when the interface is close to the locked state, for example, AB-stacking[3,6,11,13,47]. Considering many twisted interfaces in series with an average separation of $\delta$ (Supplementary Note 10), the conductance of a single slippery interface $G_{\text{slip}}$ can be calculated as $G_{\text{slip}}^{-1} - G_{\text{lock}}^{-1} = \delta(k_{o,\text{HOPG}}^{-1} - k_{o,\text{EG}}^{-1})$. Here, $G_{\text{lock}}$ is the conductance for the locked state which is about 22 GWm$^{-2}$K$^{-1}$ based on the measured $k_o$ of EG at the bulk limit and an interlayer distance of 0.34 nm. According to the cross-sectional STEM image in Fig. 1c, we estimate $\delta$ to be around 100 nm in our HOPG sample which is consistent with previous work[48]. This yields a room-temperature thermal conductance of around 600 MWm$^{-2}$K$^{-1}$ for a slippery interface which represents over 30-fold suppression compared to a locked interface. In addition, we plot in Fig. 2g the estimated $G_{\text{slip}}$ from about 180 K to 420 K by comparing $k_o$ of the 5 μm HOPG and EG mesas, which remains over 200 MWm$^{-2}$K$^{-1}$. No fine analysis of the trend was attempted considering the relatively large error bars and complications from the amorphous mesa edge.

Furthermore, to go beyond the collective effect of multiple interfaces with unknown twist angles, we devised another experimental scheme to mechanically manipulate and thermally measure one single intrinsic twisted interface (Methods and SI). To this end, we fabricated mesas with an actuation handle on HOPG (Fig. 3a). Then, by gently pushing the end of the handle with a tungsten probe (Supplementary Video 2), one of the slippery interfaces would be readily identified as the mesa top began to rotate. During the rotation, this interface remains in a slippery state with low shear resistance until it is eventually locked. To enable a direct comparison between the slippery and locked state, we first rotated the mesas until the handles on the top and bottom parts no longer overlapped (Fig. 3b). Some mesas would become locked during this process. For those still in the slippery state, we performed a second rotation to lock the interface (Fig. 3c). FDTR measurements were conducted both after the first and second rotation so that the twist angle $\theta$ between the two rotations is the only variable. The mesas locked during the first rotation were also measured twice to serve as a baseline for the experimental noise. In total, we managed to measure eight different mesas with $\theta$ ranging from 5° to 43° (Supplementary Fig. 16). The vertical position of the rotating interface generally varies from mesa to mesa, as shown by the AFM images in Fig. 3d.

Surprisingly, the FDTR phase signals for a single interface at the slippery and locked state overlap extremely well, as representatively shown in Fig. 3g. A closer inspection is provided in Fig. 3h by subtracting the signals before and after the second rotations. The phase variations induced by locking a single interface (blue dots) lie in the same range with the baseline (grey dots), suggesting that the difference between heat conduction at the slippery and locked interfacial state is beyond the experimental sensitivity. To estimate the interfacial thermal conductance at the slippery state, we simulated the FDTR phase difference via FEM. A series of conductance values ($G_{Gr}$) from 100 to 1000 MWm$^{-2}$K$^{-1}$ were assigned to the slippery state while a continuous condition was used for the locked state (Methods and Supplementary Fig. 21). Comparing the measured and calculated data, one may tentatively conclude that $G_{slip}$ is beyond 500 MWm$^{-2}$K$^{-1}$. As a further effort, we also managed to create mesas with two controlled slippery interfaces (Fig. 3e and Supplementary Video 3) and made comparisons to the state when they were both locked upon further rotations (Fig. 3f). Interestingly, some small differences from the baseline can now be observed at modulation frequencies around 1 MHz (Fig. 3g,h and Supplementary Fig. 17). The phase differences induced by simultaneously locking two interfaces generally overlap with the calculated curve for a total $G_{Gr}$ of 300 MW m$^{-2}$ K$^{-1}$, yielding $G_{slip}$ = 600 MW m$^{-2}$ K$^{-1}$ for a single interface. These values agree well with that extracted from the multi-interface measurements. Such high thermal conductance surpasses all previously measured values for layered materials by over five times[12,17,18], and approaches the highest values ever reported for any interface. This presents a fundamental challenge for probing angle-resolved thermal transport in graphite, despite our efforts in improving the experimental sensitivity.

To understand the physical mechanism underlying the twist-induced suppression of interfacial thermal transport, we first analyzed a single twisted interface via molecular dynamics simulations (Methods). Compared to pristine graphite ($\theta$ = 0°), a substantial suppression of the spectral thermal conductance is consistently observed upon an interlayer rotation of 17°, 21.8°, and 30°, especially below 10 THz (Fig. 4a), indicating reduced transmission probability of low-frequency phonons. To resolve the impact on different phonon branches, we further calculated the mode-resolved thermal conductivity for AB-stacked graphite and periodically-twisted graphite with $\theta$ = 21.8° based on density functional theory and the Peierls-Boltzmann transport equation (Methods). As shown in Fig. 4b, the reduction of $k_o$ for twisted graphite primarily arises from the transverse acoustic (TA) phonons, which was also considered by previous works[6,38]. The reduction of TA-mediated thermal conductivity is mainly attributed to the reduced group velocity based on the phonon dispersions in Fig. 4c, and also to larger weighted scattering phase space and higher scattering rates at low frequencies (Supplementary Fig. 30). These results provide direct evidence for the crucial impact of the TA phonons on the interfacial thermal transport.

In addition to the quantitative analysis, we further provide an intuitive picture for the twist-angle dependence (Fig. 4d). In particular, we highlight the correlation between thermal transport and interfacial sliding resistance as directly observed in the mesa-based experiments. First, we recall the widely accepted picture for the phenomenon of structural superlubricity, in which the energy landscape associated with the two surfaces in contact is represented by an 'egg-box' pattern of peaks and valleys[15]. When sliding at the locked state, a considerable lateral resistance appears since many energy barriers have to be crossed simultaneously. In comparison, the interface becomes rather slippery upon rotation as the energy landscapes of the top and bottom layers go out of registry which causes effective force cancellation. Likewise, when the macroscopic sliding is replaced by local vibrations, the in-plane atomic displacement in the top layer is readily passed down to the bottom layer at the locked state, leading to the efficient transmission of the transverse acoustic phonons. In contrast, force cancellation at the slippery interface prevents the bottom layer from following the vibrations of the top layer, and the phonon-mediated interfacial thermal transport is therefore impeded.

In summary, we have presented the first direct experimental study of thermal transport across the intrinsic twisted interfaces in graphite, in conjunction with simultaneous mechanical characterizations. A remarkable variation in the interfacial thermal conductance is observed which renders interlayer rotation as a promising degree of freedom for the active control of heat flow[38,49,50]. In addition, the ultrahigh thermal conductance of the slippery interfaces is of great interest for heat dissipation in emerging devices that exploit structural superlubricity[20,22–24]. Finally, we unveil a general physical picture for the twist-dependent thermal transport by highlighting the force-heat correlation, which lays the foundation for future explorations of phonon dynamics and transport in twisted vdW materials.

**Methods**

**Growth of single-crystal epitaxial graphite**

The single-crystal epitaxial graphite film (EG) was fabricated by the "isothermal dissolution–diffusion–precipitation" method[44]. First, a single-crystal nickel (Ni) foil (Zhongke Crystal Materials, Dongguan) was used as the growth substrate, and placed on an annealed solid carbon source (graphite paper, Zhongke Crystal Materials, Dongguan). Then, they were loaded into a chemical vapor deposition furnace (Tianjin Kaiheng, custom-designed) and heated to the growth temperature of 1300 °C with 800 standard cubic centimetres per minute (sccm) argon and 50 sccm hydrogen. The carbon atoms from the solid source would isothermally dissolve into and then diffuse through the Ni foil. Eventually, these carbon atoms precipitated out on the surface of the Ni foil, leading to the continuous growth of single-crystal graphene layers. This process resulted in the production of a single-crystal epitaxial graphite film after a growth duration of 10-50 hours.

**Fabrication of graphite mesas**

The graphite mesas were fabricated using a previously reported method[16,22,51]. The process is schematically shown in Supplementary Fig. 2. Freshly-cleaved highly oriented pyrolytic graphite (HOPG, ZYB grade from NT-MDT) and as-grown EG were used as the substrates. For the circular mesas, laser direct writing was used for efficiency with LOR and AZ1500 as the photoresists. For the mesas with a rotation handle, e-beam lithography was employed to achieve higher precision, with LOR and ZEP as the photoresists. After lithography, a short oxygen plasma cleaning was performed to improve adhesion and remove potential residues. Subsequently, a thick gold (Au) film (~200 nm) was evaporated on the graphite substrate with a thin titanium adhesion layer. After lift-off, graphite mesas with a height of about 1 μm were obtained through reactive ion etching ($O_2$). The thickness of the remaining metal film was around 100 nm. Careful characterizations of mesa shapes are presented in Supplementary Note 3 and Figs. 6-8.

**Microscopic characterizations**

The optical images were taken with Olympus BX51 and BX53 microscopes. The atomic force microscope (AFM) images were acquired with Cypher and MFP-3D Infinity from Asylum Research. The Raman spectra were measured using LabRAM HR Evolution, HORIBA Scientific. The scanning electron microscopy (SEM) images were obtained using Scios 2 HiVac from Thermo Fisher Scientific and Hitachi SU8220. The cross-sectional samples were prepared using Thermo Fisher Helios G4 UX. Scanning tunneling electron microscopy (STEM) was performed with FEI Titan Themis G2 300 at 300 kV.

**Experimental platform for mesa manipulation**

The mechanical shear and rotation of the graphite mesas was achieved by employing a custom-built nanopositioning platform as shown in Supplementary Fig. 24a. This setup mainly consists of a modified optical microscope (Olympus BX3M) with a 20× objective (Mitutoyo), a high-precision sample positioner, and a micro-manipulator with a piezoresistive force sensor. The sample positioner combines a three-axis piezo stage (PI P-561.3CD) for precise manipulation and a two-axis stick-slip stage (Nators, custom design) for long-distance movement. At the heart of the micro-manipulator is a tungsten probe connected to a dual-axis force sensor (Nators), both of which are mounted on a vertical stick-slip stage (Nators) for approaching the sample. The nominal resolutions of the force sensor along the normal and lateral directions are 1 μN and 0.5 μN, respectively.

**Measurement of mesa shearing resistance**

We employed the custom-built experimental platform (Supplementary Fig. 24a) with a tungsten probe (~2 μm tip radius) to measure the interlayer shear resistance of the graphite mesas (Supplementary Video 1). Typical normal and lateral force curves during the shearing process are presented in Supplementary Fig. 24c,d, with the major operations and events marked. Briefly, the process began with the probe slowly approaching the center of the mesa by using the three-axis piezo-actuated sample stage. After a gentle physical contact was made, the piezo stage continued to lift the sample until a desired normal force was achieved (loading). Then the piezo stage started moving in the same direction with the lateral force sensor at a fixed vertical position, which sheared the mesa top at a typical speed of 0.3 μm/s. After the shearing process, the piezo stage returned to its original vertical position to unload.

**Rotation of graphite mesas**

Rotation of the mesas was achieved using a tungsten probe with a tip radius of ~0.5 μm and the whole procedure is demonstrated in Supplementary Fig. 15 and Supplementary Video 2. Key to this process was the slow approaching of the tip towards the handle of the mesa and the subsequent gentle push on the side. As is well known in the field of structural superlubricity[13,16], there are typically many intrinsic twisted interfaces in HOPG which are rather slippery. Because of such slippery interfaces, the HOPG mesa was eventually separated into two parts once the torque applied by the tungsten probe became sufficiently large so that the top of the mesa began to rotate. The rotation angle can be accurately quantified by measuring the angle between the handles of the top and bottom parts. We would keep rotating the mesa top until the top and bottom handles no longer overlap with each other. Then, a set of FDTR thermal measurements would be performed. Note that during this stage, the lateral force needed to rotate the mesa was extremely small due to interfacial superlubricity.

Subsequently, the graphite mesa was rotated for a second time. This step was performed with extra care to ensure that only the top mesa handle was pushed. The force sensor was key to avoiding accidental damages during the approaching process. The second rotation stopped once the mesa top and bottom became locked together through AB-stacking[15], as indicated by the observation of no further rotation under a finite lateral force. At this locked state, another set of FDTR measurements was performed. Comparison with the data obtained after the first rotation offers insights into any potential difference between interfacial thermal transport at the slippery and locked states. Notably, the twist angle between the two rotations serves as the only variable in this comparison, since the overlapping area between the mesa top and bottom remains the same and thus does not contribute (Supplementary Fig. 18). Sometimes, the mesa would be locked during the first rotation. These mesas were used as references in Fig. 3h. In Supplementary Fig. 16, we show representative optical images of eight successfully rotated mesas after the first and second rotation. All these mesas had a nominal radius of 3 μm and a handle that was 1 μm wide.

For the experiments on mesas with two manually rotated interfaces, the procedure was similar although mesas with a slightly larger radius (3.5 μm) and handle width (1.5 μm) were used for better mechanical strength (Supplementary Video 3). After initial rotation of the mesa top, the same operation was repeated for the bottom part, separating the mesa into three parts with two slippery interfaces. This whole process is considered the first rotation, after which a FDTR measurement was performed. Subsequently, both of the two interfaces were rotated a second time into the locked state, followed by another FDTR measurement. Differences between the FDTR signals after the first and second rotation then represent contributions from the two interfaces.

**Frequency-domain thermoreflectance platform**

A frequency-domain thermoreflectance (FDTR) platform[52,53] with two continuous-wave (CW) lasers was employed, as schematically illustrated in Supplementary Fig. 9. The pump laser (405 nm wavelength) was directly modulated by the reference signal from a lock-in amplifier (Zurich Instruments HF2LI), which induced a periodic temperature variation in the sample. Meanwhile, the sample temperature was monitored by the probe laser (532 nm) based on variations of the surface reflectance as a function of the pump modulation frequency $f$ (about 50 kHz to 50 MHz). The phase and amplitude of the reflectance signal were subsequently resolved by the lock-in amplifier and fitted to a Fourier heat conduction model to obtain the thermal properties of the sample. A balanced photodetector was used to improve the signal-to-noise ratio, and sample mapping was achieved by using a piezo-actuated stage. For temperature-dependent measurements, an INSTEC cryostat was used.

For beam-offset FDTR measurements, the pump beam was swept across the probe beam by controlling a dichroic mirror, and the signal was recorded as a function of the offset distance at fixed modulation frequencies. The effective laser spot size (root-mean-square average of the pump and probe beam radii) was determined using the beam-offset signal (Supplementary Note 4), which is 2.9 μm for our 10× objective and 1.5 μm for 20×. The 10× objective was used in this work unless otherwise noted.

**Thermal measurement of graphite mesas**

The experimental procedure for FDTR thermal measurement of the graphite mesas is illustrated in Supplementary Fig. 15a. For FDTR measurement on the manually rotated graphite mesas, the 20× objective was used. To place the laser spots at the center of the mesa, a quick mapping of the surface reflectance was performed, where the mesa with Au cap can be clearly identified (Supplementary Fig. 15b). Then, the lasers were focused to the sample surface by maximizing the FDTR signal. Ten different points around the mesa center were manually selected and measured, with the average result recorded. The same procedure was also used for measuring the circular graphite mesas without the rotation handle albeit with a 10× objective.

**Finite element method for FDTR fitting**

To accurately describe the thermal transport in graphite mesas, the finite element method was used to calculate the phase response. For circular mesas, an axisymmetric two-dimensional model was employed for efficient fitting of the out-of-plane thermal conductivity, as shown in Supplementary Fig. 20a. The model was constructed and solved by using the PDE interface of COMSOL Multiphysics. The heat conduction equation is given in the frequency domain as[54]

$$k_i \frac{\partial^2 T_f}{\partial r^2} + \frac{k_i}{r}\frac{\partial T_f}{\partial r} + k_o \frac{\partial^2 T_f}{\partial z^2} = 2\pi j f C T_f. \qquad (1)$$

Here, $T_f(f,r)$ is the Fourier transform of the temperature, $j$ is the imaginary unit, $C$ is the heat capacity, and $k_i$ and $k_o$ denote the in-plane and out-of-plane thermal conductivity, respectively. For the boundary conditions, a constant-temperature boundary was employed on the side and bottom surface of the substrate; a continuous boundary condition was used between the graphite mesa and the substrate; a thermal boundary conductance $G_{\text{metal}}$ was added between the metal layer and the graphite mesa; and a heat flux $q$ with a Gaussian distribution was applied to the top surface of the metal cap. The simulated FDTR phase signal was extracted as $\varphi = \text{phase}(\overline{T_f})$, with $\overline{T_f}$ given by an integral over the top surface:

$$\overline{T_f} = \int_0^{r_{\text{mesa}}} \exp\left(-\frac{2r^2}{w_{\text{probe}}^2}\right) 2\pi r T_f(f,r) dr. \qquad (2)$$

Here, $w_{\text{probe}}$ and $r_{\text{mesa}}$ are the probe and mesa radii, respectively. For graphite mesas with the rotation handle, a three-dimensional model was built, as shown in Supplementary Fig. 21a. An interfacial thermal conductance $G_{Gr}$ was introduced for the slippery state, while a continuous condition was used for the locked state. The governing equation and other conditions were kept the same.

**Molecular dynamics simulations**

We employed nonequilibrium molecular dynamics simulations to calculate the interfacial thermal conductance and analyze the transport spectrum using the Large-scale Atomic/Molecular Massively Parallel Simulator (LAMMPS) package[55]. As shown in Supplementary Fig. 26, a 20-layer model was constructed in which the top 10 layers were twisted by an angle $\theta$ with respect to the bottom 10 layers. All the layers within the top and bottom parts remain AB-stacked. The lateral sizes of the simulation domains were approximately 40 × 40 Å$^2$. Periodic boundary conditions were then applied to all directions. The intralayer interactions were described by the optimized Tersoff potential[56]. For the interlayer van der Waals (vdW) force, the Lennard-Jones potential was employed with a dispersion energy of 4.6 meV and a zero-potential-energy distance parameter ($\sigma$) of 0.3276 nm, respectively[11,57]. The cut-off distance was set to $3\sigma$.

In the MD simulations, the system energy was first minimized using the conjugate gradient method. The obtained atomic structure was further relaxed in the isothermal-isobaric (NPT) ensemble at 300 K and zero pressure for 1.5 ns. We then switched to the canonical (NVT) ensemble and maintained for another 2 ns. Subsequently, the system was kept in the microcanonical (NVE) ensemble for 5 ns. During this stage, Langevin thermostats were applied to the two layers adjacent to the fixed layers, setting the temperatures to 350 K and 250 K to serve as the heat source and sink, respectively (Supplementary Fig. 26). A time step of 0.5 fs was used throughout these stages. The heat flux was recorded after a stable temperature distribution was generated (Supplementary Fig. 27). The interfacial thermal conductance was then calculated by

$$G = \frac{q}{A\Delta T}, \tag{8}$$

where $A$ is the cross-sectional area and $q$ is the heat flux which was calculated by monitoring the energy change in the heat source and sink. The spectral heat conductance was calculated following previous works[11,58,59].

**Density functional theory calculation**

For twisted graphite, we employed a periodically stacked system composed of twisted bilayer graphene with $\theta = 21.8°$ (Supplementary Fig. 28), which has 28 atoms in the unit cell and is the smallest in all the twisted graphene configurations[60]. However, the computational cost remained huge and prevented the calculation of the third-order interatomic force constants (IFCs) using traditional methods[61], and we therefore turned to an approach based on parameter fitting and advanced machine learning algorithms using the hiPhive package[62]. Four-phonon scattering was ignored considering its limited impact and even higher cost. For direct comparison, an AB-stacked graphite system was also constructed.

In the density functional theory (DFT) calculations, we employed the Perdew-Burke-Ernzerhof exchange-correlation functional[63] and projector-augmented wave method[64,65] as implemented in the Vienna Ab initio Simulation Package (VASP)[64,66]. The optB86b-vdW functional[67] was used to describe the interlayer vdW interactions. The cutoff energy and convergence threshold were chosen as 800 eV and $10^{-8}$ eV, respectively. We used 15×15×5 and 5×5×5 **k**-meshes for the AB-stacked and twisted graphite, respectively, for geometry optimization. The second-order IFCs were calculated using density functional perturbation theory and postprocessed in the Phonopy package[68] with 6×6×2 and 2×2×2 supercells for the AB-stacked and twisted case. For the third-order IFCs, the cutoffs in the hiPhive package were set to 6 Å and 4 Å for pairs and triplets, respectively.

The thermal conductivity was obtained by solving the Peierls-Boltzmann transport equation for phonon transport with the ShengBTE code[61], including three-phonon and isotope-phonon scattering processes. We used 10×10×10 and 30×30×10 **q**-meshes for twisted and AB-stacked graphite, respectively. In addition, the broadening factor was set to unity to ensure a full convergence.

**Data availability:** All data needed to evaluate the conclusions in the paper are present in the main text or the supplementary information.

**Method references**

**Acknowledgments**

This work was financially supported by the National Natural Science Foundation of China (Nos. 52076002, 11890671, 51961145304, 52025023, and 52076089), Ministry of Science and Technology of China (No. 2022YFA1203100), and Guangdong Major Project of Basic and Applied Basic Research (No. 2021B0301030002), and the Tsien Excellence in Engineering program. B.S. and K.L. acknowledge support from the New Cornerstone Science Foundation through the XPLORER PRIZE. We thank the Molecular Materials and Nanofabrication Laboratory of Peking University and Shenzhen Key Laboratory of Superlubricity Technology (ZDSYS20230626091 701002) for supporting our experiments. We appreciate the High-performance Computing Platform of Peking University and Huazhong University of Science and Technology for supporting our simulations. F.Y. appreciates Li Chen and Zipei Tan for help with the experimental platform.

**Author contributions**

B.S., Q.Z., and F.Y. conceived the idea and designed the research. Z.Z. and M.D. grew the epitaxial graphite. F.Y. designed the graphite mesas while X.H. and Y.H. did the microfabrication. F.Y. performed the mechanical measurements and analyses. F.Y., W.Y., and Y.W. conducted the thermal measurements and analyses. W.Z. and N.L. performed DFT calculations. J.Z. performed the MD simulations. F.Y., X.H., Y.H., Z.Z., and Q.G. performed the optical, AFM, Raman, SEM, and STEM characterizations. B.S., Q.Z., K.L., and T.-H.L. supervised the work. F.Y. and B.S. prepared the manuscript with input from all authors.

**Competing interests**

Authors declare no competing interests.

**Additional information**

**Supplementary information is available for this paper.**

**Correspondence and requests for materials** should be addressed to Bai Song, Quanshui Zheng, Kaihui Liu, and Te-Huan Liu.

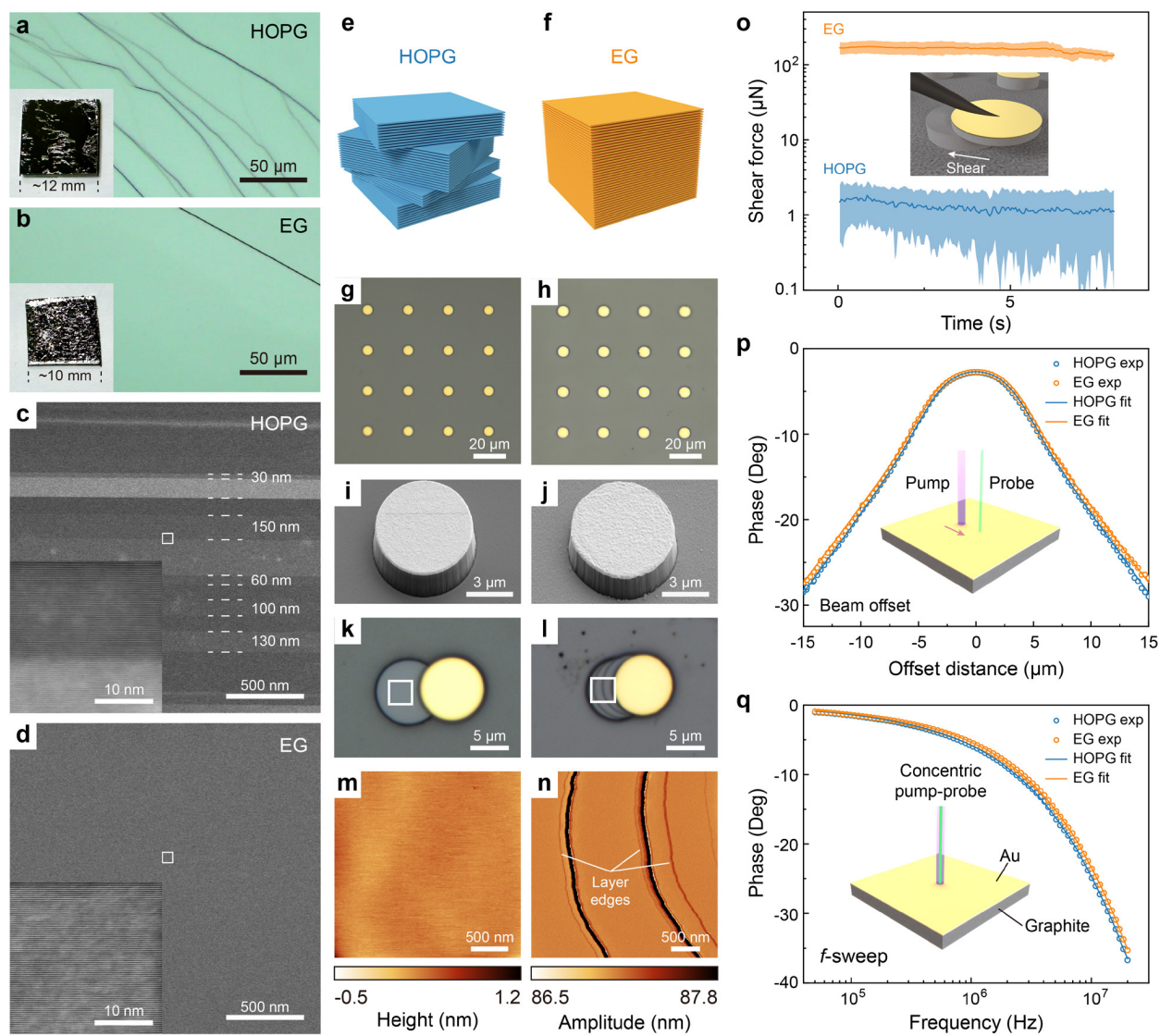

**Fig. 1 | Structural, mechanical, and preliminary thermal characterizations of graphite. a**, **c**, **e**, **g**, **i**, **k**, and **m** are for HOPG, while **b**, **d**, **f**, **h**, **j**, **l**, and **n** are for EG. **a** and **b**, Optical images showing representative samples (insets) and surfaces. **c** and **d**, Cross-sectional STEM images with the intrinsic twisted interfaces in HOPG labeled. Insets show atomically resolved structures with the white squares. **e** and **f**, Schematics of the layered structures. **g** and **h**, Optical images of graphite mesa arrays. **i** and **j**, SEM images of individual mesas. **k** and **l**, Optical images of representative sheared mesas. **m** and **n**, AFM mapping within the white boxes in **k** and **l**, respectively. Here, **m** shows the surface topography while the amplitude signal is used in **n** to better resolve the layer edges. **o**, Shear forces for mesas on HOPG and EG. **p** and **q**, FDTR phases from room-temperature beam-offset and $f$-sweep measurements together with the fitted curves, respectively.

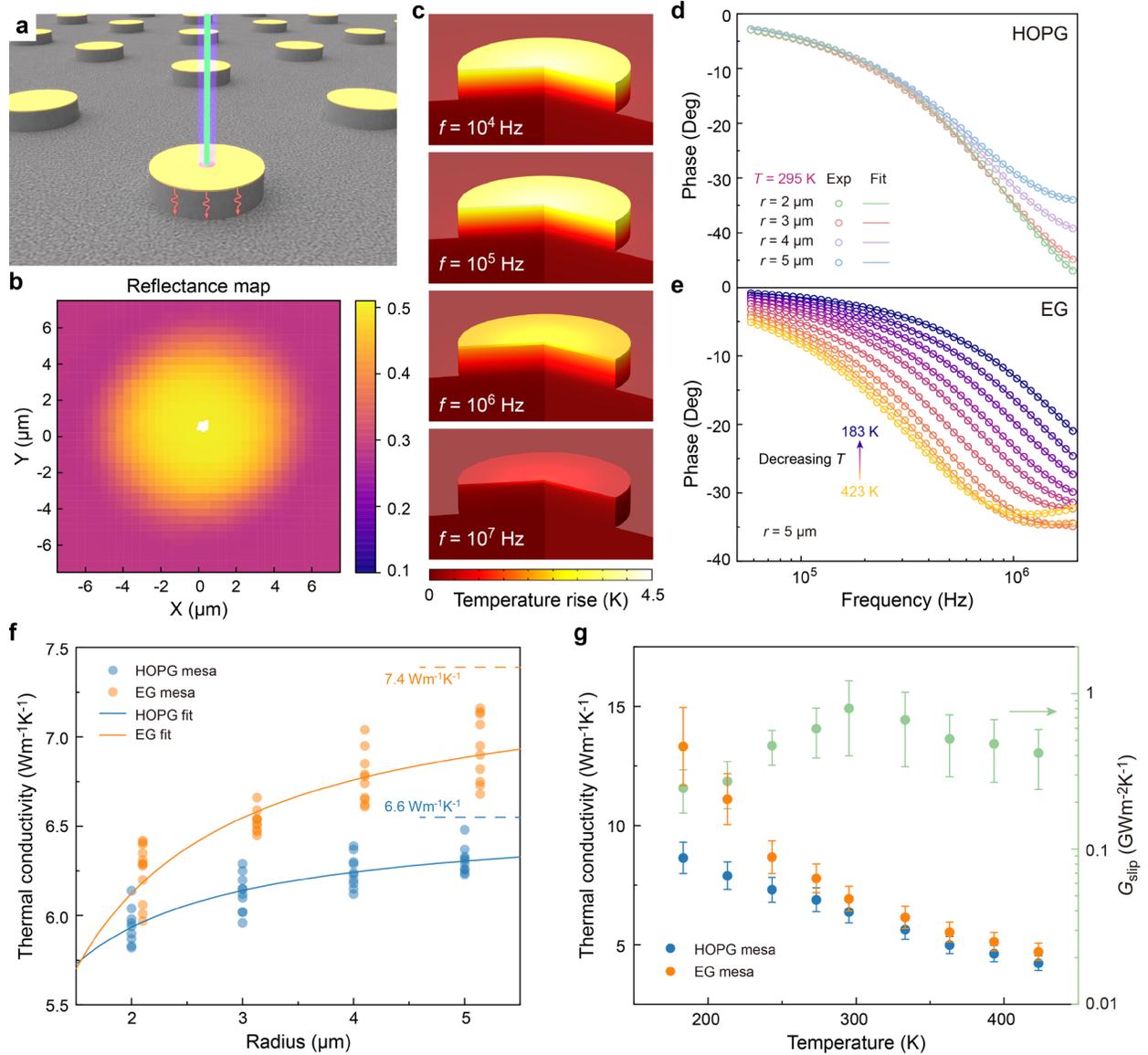

**Fig. 2 | Thermal measurement of multiple intrinsic twisted interfaces. a**, Schematic of mesa-based FDTR measurement. **b**, Reflectance mapping of the mesa. Ten white dots at the center mark the locations for collecting single-point FDTR phase signals. **c**, FEM simulations of the temperature distributions at different modulation frequencies. **d**, Representative FDTR phases together with the fitted curves for HOPG mesas with different radii. **e**, Representative FDTR phases together with the fitted curves for EG mesas at different temperatures. **f**, Measured and modeled out-of-plane thermal conductivity as a function of mesa radius for both HOPG and EG. The values at the large-radius limit are labeled. **g**, Thermal conductivity with varying temperature for mesas with a nominal radius of 5 μm. Right axis shows the estimated thermal conductance of slippery interfaces.

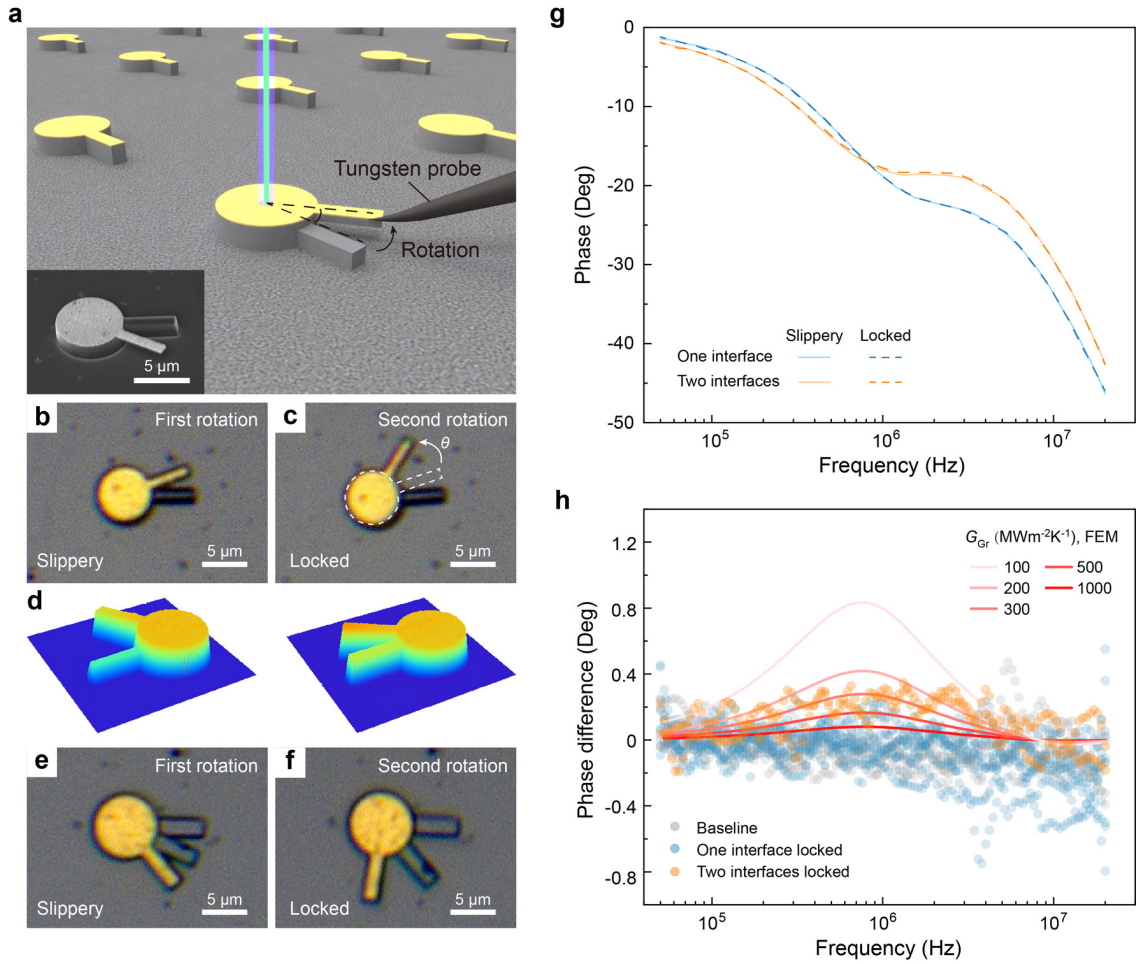

**Fig. 3 | Thermal measurement of intrinsic interfaces after manual rotation. a**, Schematic for mesa rotation and FDTR measurement. Inset: SEM image of a mesa after rotation. **b** and **c**, Optical images for the same representative mesa at the slippery state after the first rotation and locked after the second rotation, respectively. **d**, AFM images of two rotated mesas at the locked state. **e** and **f**, Optical images showing the same mesa with two interfaces simultaneously at the slippery state and the locked state, respectively. **g**, Typical FDTR phase signals at the slippery and locked states for mesas with one and two manually rotated interfaces. **h**, FDTR phase difference between the first and second rotation, together with FEM simulated results assuming a series of thermal conductance values for the slippery interface. The light grey, blue, and orange dots consist of data from six, eight, and two different mesas, respectively.

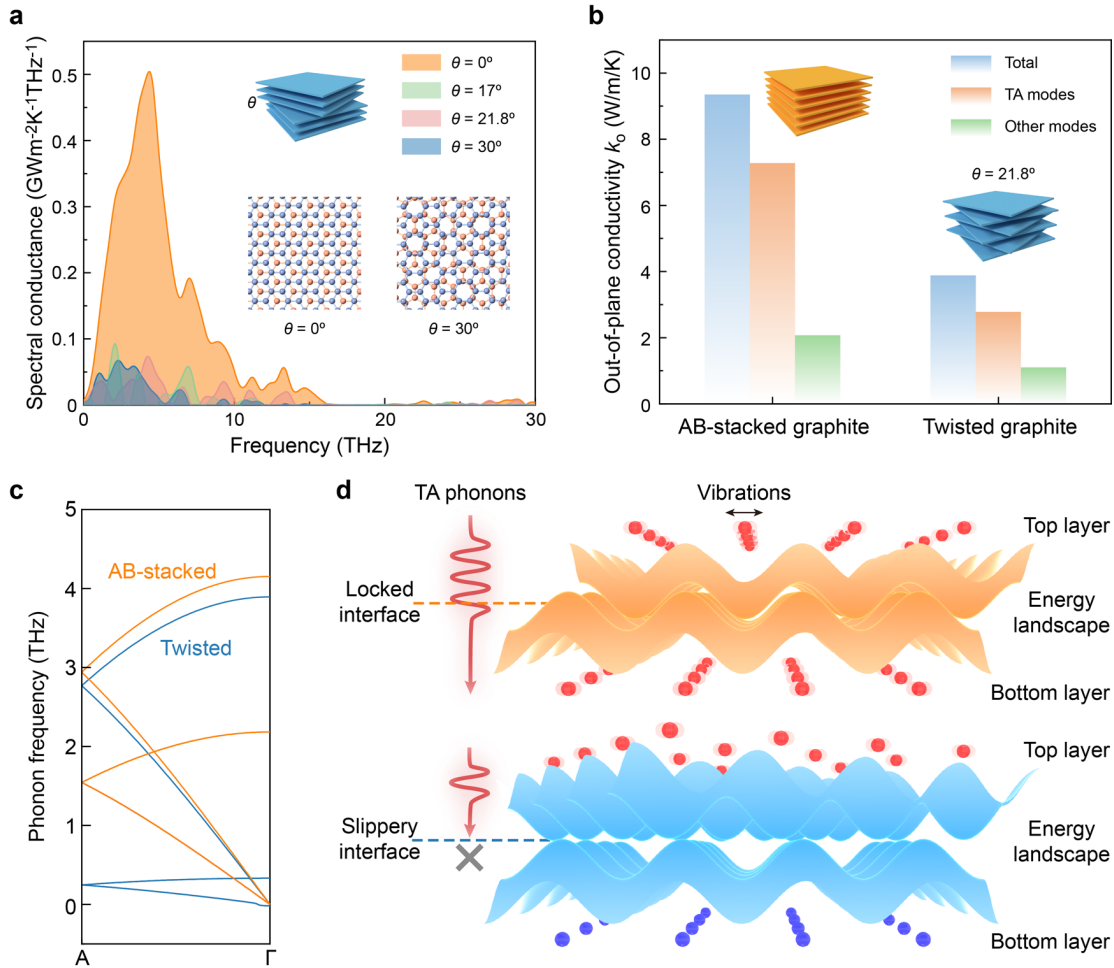

**Fig. 4 | Mechanism of twist dependence and force-heat correlation. a**, Spectral interfacial thermal conductance at different twist angles from MD simulations. Insets: schematic of the twisted graphite model together with atomic structures for the cases of AB-stacking ($\theta = 0°$) and 30° twist. **b**, Mode-resolved out-of-plane thermal conductivity for AB-stacked and twisted graphite. Inset: schematics of single-crystal and twisted graphite employed in DFT calculations. **c**, Out-of-plane phonon dispersions for AB-stacked and twisted graphite calculated by DFT. **d**, Schematic illustration of phonon transmission across slippery and locked interfaces.